\documentclass[acmtog]{acmart}
\setcopyright{rightsretained}
\copyrightyear{2026}
\acmYear{2026}
\acmDOI{10.1145/3592405}

\acmSubmissionID{1365}

\acmJournal{TOG}
\acmVolume{42}
\acmNumber{4}
\acmArticle{1}
\acmYear{2026}
\acmMonth{8}

\usepackage{booktabs}

\graphicspath{{./images/}}

\usepackage{wrapfig}

\usepackage{comment}

\usepackage{soul}

\usepackage[linesnumbered]{algorithm2e}

\usepackage{xargs} % Use more than one optional parameter in a new commands

\newif\ifsubmit
\submittrue
\ifsubmit

    \usepackage[colorinlistoftodos,prependcaption,textsize=tiny,color=todo,tickmarkheight=0.1cm]{todonotes}
    \setuptodonotes{noinline}

    \newcommand{\TODO}[1]{}

    \newcommand{\revbegin}{}
    \newcommand{\revend}{}

    \newcommand{\deleted}[1]{}

    \newcommand{\sayit}[3]{}
    
    \newcommand{\yotam}[2][1=]{}
    \newcommand{\ted}[2][1=]{}
    \newcommand{\jason}[2][1=]{}
    \newcommand{\jianchao}[2][1=]{}
    \newcommand{\jose}[2][1=]{}
\else
    \definecolor{todo}{rgb}{0.99,0.59,0.55}
    \newcommand{\TODO}[1]{\textcolor{red}{\textbf{****** #1 ******}}}

    \newcommand{\revbegin}[1]{\color{blue}} % \revbegin
    \newcommand{\revend}[1]{\color{black}} % \revend

    \newcommand{\deleted}[1]{\st{#1}}

    \newcommand{\sayit}[3]{{\small\protect\colorlet{col}{#2}\color{col}\colorbox{col!15}{\textsc{#1:}} #3}}
    
    \definecolor{cyan3}{HTML}{C99E10} %
    \usepackage[colorinlistoftodos,prependcaption,textsize=tiny,color=todo,tickmarkheight=0.1cm]{todonotes}
    \setuptodonotes{noinline}
    \definecolor{lime}{rgb}{0.53,0.78,0.27}
    \newcommandx{\yotam}[2][1=]{\todo[linecolor=lime,backgroundcolor=lime!25,bordercolor=lime,#1]{\colorbox{lime!50}{\textsc{Yotam:}} #2}}
    \newcommandx{\ted}[2][1=]{\todo[linecolor=teal,backgroundcolor=teal!25,bordercolor=teal,#1]{\colorbox{teal!50}{\textsc{Ted:}} #2}}
    \newcommandx{\jason}[2][1=]{\todo[linecolor=orange,backgroundcolor=orange!25,bordercolor=orange,#1]{\colorbox{orange!50}{\textsc{Jason:}} #2}}
    \newcommandx{\jianchao}[2][1=]{\todo[linecolor=blue,backgroundcolor=blue!25,bordercolor=blue,#1]{\colorbox{blue!50}{\textsc{Jianchao:}} #2}}
    \newcommandx{\jose}[2][1=]{\todo[linecolor=magenta,backgroundcolor=magenta!25,bordercolor=magenta,#1]{\colorbox{magenta!50}{\textsc{Jose:}} #2}}
    
\fi

\usepackage{xspace}

\definecolor{quote-gray}{gray}{0.3}

\renewcommand{\paragraph}[1]{\emph{#1}}

\usepackage{dblfloatfix}

\usepackage{etoolbox}
\AtBeginEnvironment{equation}{\vspace{-.75ex}}
\AtEndEnvironment{equation}{\vspace{-.75ex}}

\begin{document}

\newcommand{\system}{\emph{ColorGradedGaussians}\xspace}
\title{\system: Palette-Based Color Grading for 3D Gaussian Splatting via View-Space Sparse Decomposition}
% Palette-Based Color, Curve, and Pixelwise Color Grading for 3D Gaussian Splatting with View-space Sparse Decomposition

\author{Cheng-Kang Ted Chao}
\email{cchao8@gmu.edu}
\affiliation{%
  \institution{George Mason University}
  \country{USA}
}

\author{Yotam Gingold}
\email{ygingold@gmu.edu}
\affiliation{%
  \institution{George Mason University}
  \country{USA}
}

\renewcommand{\shortauthors}{Cheng-Kang Ted Chao and Yotam Gingold}

\begin{teaserfigure}%
\centering
\includegraphics[width=0.95\textwidth]{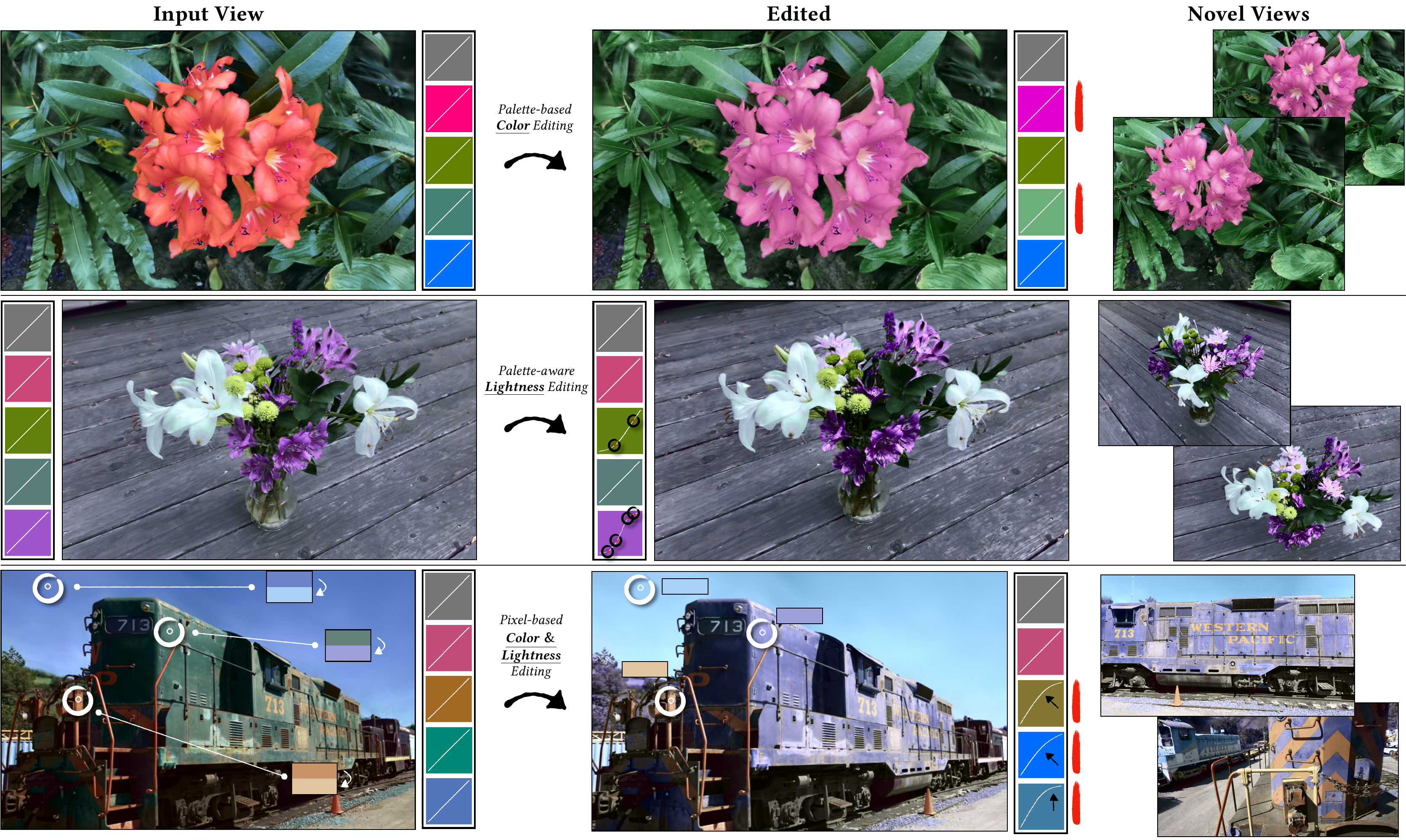}
\caption{\system enables real-time palette-based color grading for 3D Gaussian Splatting. Our method supports three editing modes: \textbf{(Top)} Palette editing: changing two palette colors makes the flowers pink and the leaves greener. \textbf{(Middle)} Per-palette tone curves: S-curves on specific colors (yellow-green and purple) add contrast to leaves and purple flowers independently without affecting white flowers. \textbf{(Bottom)} Pixel-level constraints: three color targets (sky, train, stairs) optimize the palette and tone curves to achieve a winter atmosphere with a purple train. All edits are real-time and propagate consistently to novel views (right).
Our method achieves superior sparsity by splatting weights instead of colors and optimizing for view-space sparsity.
% \yot{I simplified the first two sentences and described view-space sparsity above. Lost are: ``view-space palette decomposition'' and ``operates in CIE LAB color space''.}
Datasets (top to bottom): \textit{Flower} from LLFF dataset \cite{mildenhall2019local}; \textit{Vasedesk} from Mip-NeRF 360 \cite{barron2022mip}; \textit{Train} from Tanks\&Temple \cite{knapitsch2017tanks}.}
\label{fig:teaser}
\Description{}
\end{teaserfigure}

\begin{abstract}
Professional color editing requires precise control over both color (hue and saturation) and lightness, ideally through separate, independent controls. We present a real-time interactive color editing framework for 3D Gaussian Splatting (3DGS) that enables palette-based recoloring, per-palette tone curves for color-aware lightness adjustment, and accurate pixel-level constraints---capabilities unavailable in prior palette-based 3DGS methods. Existing approaches decompose colors at the primitive level, optimizing per-Gaussian palette weights before splatting. However, sparse primitive-level weights do not guarantee sparse pixel-level decompositions after alpha-blending, causing palette edits to affect unintended regions and degrading editing quality. We address this through view-space palette decomposition, splatting weights instead of colors to optimize the observable appearance of the scene. We introduce a geometric loss using inverse barycentric coordinates to enforce consistent sparsity patterns, ensuring similar colors share similar decompositions. Our approach achieves superior editing quality compared to primitive-space methods, enabling professional color grading workflows for 3DGS scenes with real-time interaction.
\end{abstract}




\maketitle

\section{Introduction}
Color grading for 3D scenes is central to modern content creation, from visual effects and film production~\cite{debevec2008rendering} to virtual production and augmented reality~\cite{jeffrey2020ves}. 
Beyond global adjustments, artists need selective control over specific color ranges, with edits that remain photorealistic and geometrically consistent across viewpoints. 
In 2D, palette-based editing~\cite{chang2015palette, Tan:RGB2016, tan2018efficient} offers an intuitive workflow by expressing each pixel as a mixture of a small set of palette colors and propagating edits by keeping the mixing weights fixed. 
This paradigm has been improved through better palette extraction and sparser, more localized decompositions~\cite{aksoy2017unmixing, wang2019improved, chao2023locopalettes, zhang2025semantic}. Professional workflows benefit from independent lightness control per color range, e.g., via palette-aware tone curves~\cite{chao2023colorfulcurves}. 
Extending these capabilities to 3D scenes requires not only multi-view consistency, but also palette decompositions that remain sparse and editable in the rendered images users observe and edit.

Recent reconstruction methods such as NeRF~\cite{mildenhall2021nerf} and 3D Gaussian Splatting (3DGS)~\cite{kerbl20233d} enable photorealistic novel-view synthesis.
% motivating palette-based editing in 3D. 
Several methods adapt palette-based editing to NeRF~\cite{kuang2023palettenerf, gong2023recolornerf, wu2022palettenerf} and 3DGS~\cite{PaletteGaussian, RecolorGaussian}. The 3DGS methods decompose each Gaussian's diffuse color into weighted combinations of global palette colors.
% However, existing palette-based 3DGS methods~\cite{PaletteGaussian, RecolorGaussian} suffer from two critical limitations. First, they optimize per-Gaussian palette weights that are then splatted to form rendered images.
Because splatting blends multiple Gaussians,
% through non-linear operations \yotam{does the non-linearity affect the lack of sparsity?} (Gaussian falloff and depth-dependent transmittance),
sparse per-Gaussian weights do not, in general, lead to view-space (per-pixel) sparsity. This causes palette edits to affect unintended regions (Fig.~\ref{fig:sparsity-compare-flip}). Palette-based NeRF methods~\cite{kuang2023palettenerf} face similar challenges in ensuring view-space sparsity and additionally require expensive preprocessing for weight guidance. Moreover, these methods lack color-aware tone curves for independent lightness adjustment per palette color. This is a fundamental capability in professional color grading~\cite{chao2023colorfulcurves}. Finally, directly editing pixel colors requires joint re-optimization of palette and view-dependent representations, preventing real-time interaction (Fig.~\ref{fig:pixel-editing-compare}).

\begin{figure}
\includegraphics[width=\linewidth,scale=1]{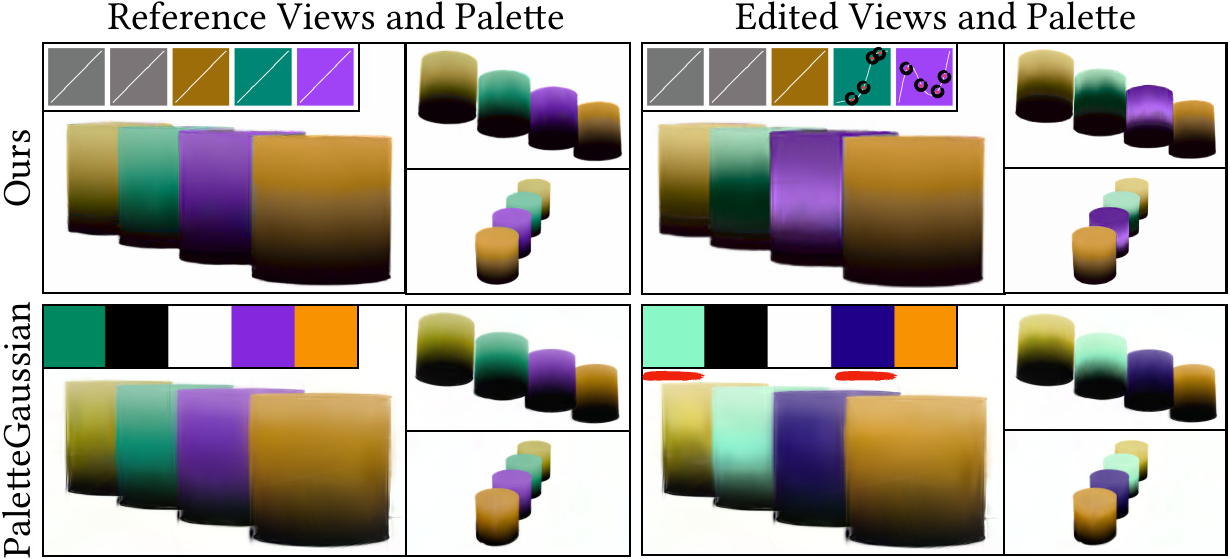}
\caption{Our method (top) enables independent lightness control per palette color on 3DGS via tone curves. We can independently add contrast to green (S-curve) and adjust purple's tonal range (black control points). PaletteGaussian~\cite{PaletteGaussian}'s linear RGB mixing (bottom) lacks this capability. Palette color changes affect hue, saturation, and lightness simultaneously, preventing independent tonal control.
%% Yotam: Do we really need to credit ourselves and BlenderNeRF? We can include it in the bibliography with \nocite{}. UPDATE: \nocite isn't working.
% (Dataset via BlenderNeRF \cite{Raafat_BlenderNeRF_2024}.)
% \nocite{Raafat_BlenderNeRF_2024}
%\TODO{mention our palette is visualized in L=50.}
% \TODO{Trim whitespace around the images tighter.}
}
\label{fig:palette-aware-lightnes-compare}
\end{figure}

We present \system, a real-time color grading framework for 3DGS that resolves these limitations by performing palette decomposition in view space.
% \revised{and provides per-pixel color and per-palette lightness controls}.
Our key idea is to splat palette weights rather than colors, producing per-pixel mixing weights in the rendered view where sparsity is perceived and edits are specified. 
This shift enables effective view-space sparsity regularization and supports three complementary controls: direct palette editing, per-palette tone curves for color-aware lightness adjustment, and accurate pixel-level constraints from any rendered view, all with real-time updates (Fig.~\ref{fig:teaser}). 
Achieving sparsity alone is insufficient for high-quality editing; similar colors must also share consistent sparsity patterns to avoid unpredictable edit propagation (Fig.~\ref{fig:geometric-sparsity-ablation}). 
We therefore introduce a geometric consistency loss based on inverse barycentric coordinates that encourages perceptually similar colors to receive similar decompositions. 
We evaluate \system on multiple datasets and show that it provides editing capabilities unavailable in prior palette-based 3DGS methods, including per-palette tone curves and accurate pixel-level constraints, while also improving edit locality and interactivity.

\section{Related Work}
\paragraph{Appearance editing in neural scene representations}
Neural scene representations such as NeRF~\cite{mildenhall2021nerf} and 3D Gaussian Splatting (3DGS)~\cite{kerbl20233d} have enabled a wide range of appearance editing techniques by explicitly modeling view-dependent radiance. 
Text-driven editing methods~\cite{haque2023instruct, zhuang2023dreameditor, chen2024gaussianeditor, kamata2023instruct, bao2023sine} and neural style transfer approaches~\cite{zhang2022arf, gu2021stylenerf, liu2024stylegaussian, huang2022stylizednerf, fan2022unified, zhang2025stylizedgs} allow semantic or artistic modifications but lack direct or precise color control and often struggle with multi-view consistency, making them unsuitable for professional color grading workflows. 
Relighting and tone mapping methods~\cite{gao2024relightable, cui2025luminance, liu2025gausshdr} focus on photometric consistency through physically based rendering or luminance-domain processing, but do not offer palette-based abstractions or selective color control. 
To enable precise color editing, several works introduced palette-based representations for neural scenes. \citet{tojo2022recolorable} extends posterization-style editing~\cite{chao2021posterchild} to NeRFs for synthetic scenes, while PaletteNeRF~\cite{kuang2023palettenerf} and RecolorNeRF~\cite{gong2023recolornerf} support real-world scenes via palette decomposition of volumetric points, at the cost of expensive preprocessing and slow rendering. 
More recent palette-based 3DGS methods~\cite{PaletteGaussian, RecolorGaussian} achieve real-time performance by decomposing each Gaussian's color into mixtures of global palette colors. However, their primitive-based decomposition assigns weights per Gaussian that are later combined through
% non-linear \yotam{again, is non-linear relevant or just the combination?}
splatting. Sparsity at the primitive level does not translate into sparse or localized edits in the rendered image, manifesting as color bleeding (Fig.~\ref{fig:sparsity-compare-flip}). 
Our work directly addresses this limitation by shifting palette decomposition to view space, where edits are observed and enforced.

\paragraph{Palette-based image recoloring}
Palette-based image recoloring enables intuitive color control by representing each pixel as a mixture of a small set of representative palette colors.
\citet{chang2015palette} introduced this paradigm through palette extraction via color clustering and recoloring using radial-basis-function-weighted color deformation.
Subsequent work formulated palette extraction and decomposition geometrically in color space. \citet{Tan:RGB2016} proposed convex-hull-based decomposition into translucent color layers. \citet{tan2018efficient} extended this framework to compute spatially coherent additive mixing weights in RGBXY-space, enabling recoloring and color harmonization~\cite{tan2025palette}.
\citet{wang2019improved} introduced an optimization-based approach to create more representative palette polyhedra.
\citet{sun2023building} introduced a coarse-to-fine convex hull construction to improve palette quality.
Independent of palette extraction, \citet{chao2023colorfulcurves} introduced palette-aware tone curves to support independent lightness and contrast control per palette color and image-space color constraints.
Extensions to video recoloring compute temporally coherent RGBT palettes~\cite{Du:2021:VRS}, with later work accelerating palette evolution via Bézier interpolation between frames~\cite{du2025fast}.
A fundamental limitation of palette-based recoloring is its difficulty in selectively editing objects that share similar colors. This has been addressed by incorporating segmentation~\cite{chao2023locopalettes} and by extracting semantic palettes in high-dimensional feature spaces~\cite{du2024palette, zhang2025semantic}.
Alternative unmixing approaches~\cite{aksoy2016interactive, aksoy2017unmixing, koyama2018decomposing} instead formulate recoloring as energy minimization to produce sparse layers of nearly homogeneous colors.
Crucially, all these methods operate directly in image space, where palette weights are defined, regularized, and edited at the pixel level. This assumption breaks down when extending palette-based editing to 3D scenes, yet motivates our view-space formulation.

\section{Background and Motivation}
We begin by reviewing the mathematical formulation of 3D Gaussian Splatting (3DGS)~\cite{kerbl20233d}, which forms the foundation of our work, before discussing the limitations of state-of-the-art palette-based decomposition approaches~\cite{PaletteGaussian, RecolorGaussian} that motivate our view-space formulation.

\paragraph{Background}
\label{sec:background}
3DGS has emerged as a powerful real-time rendering technique that represents scenes as collections of 3D Gaussians. Each Gaussian $i$ is parameterized by a center position $\boldsymbol{\mu}_i \in \mathbb{R}^3$, covariance matrix $\boldsymbol{\Sigma}_i \in \mathbb{R}^{3 \times 3}$, opacity $\alpha_i \in [0,1]$, and view-dependent color $\mathbf{c}_i \in \mathbb{R}^{3}$. For a given camera position $(\theta, \phi)$, the color is typically encoded using spherical harmonics (SH) coefficients $f_{l,m} \in \mathbb{R}$, where $l$ denotes the SH band and $m \in \{-l, \ldots, l\}$ denotes the order within each band, to capture view-dependent appearance:
% \begin{equation} \label{eq:3dgs_view_color}
$
\mathbf{c}_i(\theta, \phi) = \sum_{l=0}^{L} \sum_{m=-l}^{l} f_{l,m} Y_l^m(\theta, \phi),
$
% \end{equation}
where $Y_l^m(\theta, \phi)$ are the spherical harmonic basis functions and $L$ is the maximum band.
To render an image from a given viewpoint, each 3D Gaussian is projected onto the 2D image plane~\cite{zwicker2001ewa}. The final pixel color is computed by $\alpha$-blending the contributions of all overlapping Gaussians in front-to-back depth order:
\begin{equation}\label{eq:3dgs_rendering}
\mathbf{C}(\mathbf{x}; \theta, \phi) = \sum_{i \in \mathcal{N}_\mathbf{x}} \mathbf{c}_i(\theta, \phi) \alpha_i' \prod_{j=1}^{i-1}(1 - \alpha_j')
\end{equation}
where $\mathcal{N}_\mathbf{x}$ is the set of Gaussians affecting pixel $\mathbf{x}$ sorted by depth, $\mathbf{c}_i(\theta, \phi)$ is the view-dependent color of Gaussian $i$, and 
\begin{equation}
% $
\alpha_i' = \alpha_i \exp\left(-\frac{1}{2}(\mathbf{x} - \boldsymbol{\mu}_i')^T {\boldsymbol{\Sigma}_i'}^{-1} (\mathbf{x} - \boldsymbol{\mu}_i')\right)
\end{equation}
% $
is the Gaussian's opacity modulated by its 2D spatial footprint.
The Gaussian parameters are optimized from multi-view images using stochastic gradient descent with adaptive density control that splits and prunes Gaussians during training~\cite{kerbl20233d}.

\begin{figure*}[!t]
\includegraphics[width=\linewidth,scale=1]{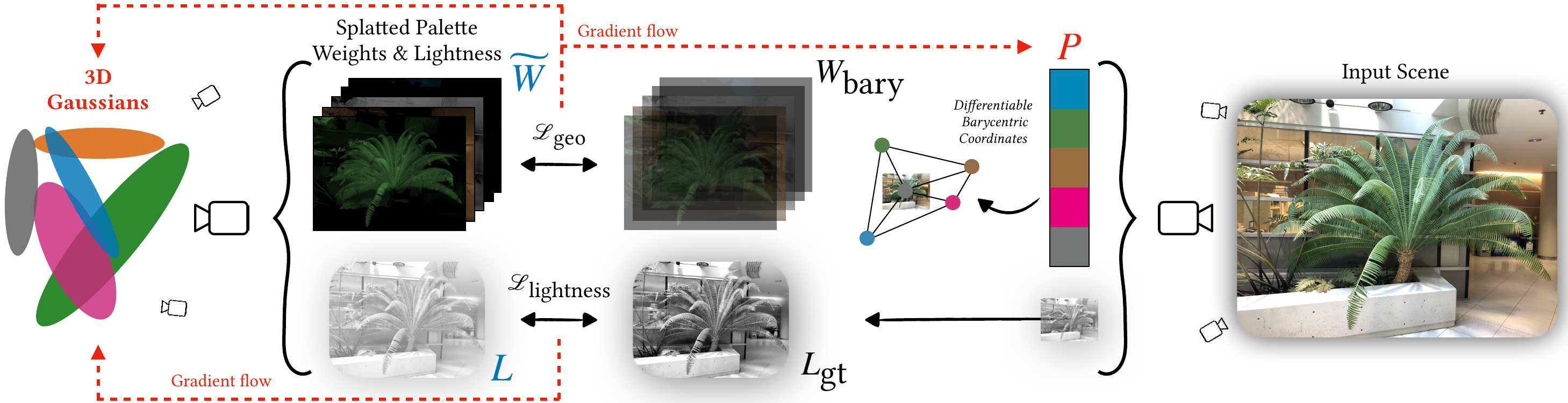}
\caption{\textbf{Training Pipeline.} Our method trains 3D Gaussians (left, red label) that splat palette weights $\widetilde{\mathbf{W}}$ and lightness $L$ in CIELAB color space, with the palette $\mathbf{P}$ (right, red label) optimized in \textit{ab}-space only.
% Trained parameters (3D Gaussians, palette $\mathbf{P}$) are labeled in red.
The splatted weights are supervised by computing differentiable inverse barycentric coordinates from the input images with respect to the current palette state, yielding geometric target weights $\mathbf{W}_{\text{bary}}$. The geometric loss $\mathcal{L}_{\text{geo}}$ (Eq.~\ref{eq:geometric_loss}) enforces consistency between $\widetilde{\mathbf{W}}$ and $\mathbf{W}_{\text{bary}}$, while the lightness loss $\mathcal{L}_{\text{lightness}}$ (Eq.~\ref{eq:lightness_loss}) directly supervises the splatted lightness $L$ against ground truth $L_{\text{gt}}$. Red dashed arrows indicate gradient flow, with losses backpropagating to both the 3D Gaussian parameters and the palette $\mathbf{P}$. The final rendered image is obtained by multiplying the splatted weights with the palette and concatenating with the lightness channel to form complete Lab images. Additional regularization terms (not shown) include palette regularization (Sec.~\ref{sec:palette_reg}) and sparsity regularization on the weights (Eq.~\ref{eq:sparsity_loss}). (\textit{Fern} from the LLFF dataset \cite{mildenhall2019local}.)}
\label{fig:pipeline}
\end{figure*}

\paragraph{Palette-based 3DGS Recoloring}
\label{sec:motivation}
Recent work~\cite{PaletteGaussian, RecolorGaussian} has extended 3DGS to support palette-based color editing by decomposing each Gaussian's color into a weighted combination of palette colors. Specifically, instead of directly computing view-dependent colors $\mathbf{c}_i$ for each Gaussian $i$, these methods perform color decomposition at the Gaussian level. Each Gaussian's diffuse color is represented as a linear combination of a trainable global palette $\mathbf{P} = [\mathbf{p}_1, \mathbf{p}_2, \ldots, \mathbf{p}_K] \in \mathbb{R}^{K \times 3}$ shared across the entire scene:
\begin{equation} \label{eq:palettegaussian_diffuse}
\mathbf{c}_i^d = \sum_{k=1}^{K} w_{i,k} \mathbf{p}_k = \mathbf{w}_i^{\top} \mathbf{P}
\end{equation}
where $\mathbf{w}_i = [w_{i,1}, \ldots, w_{i,K}]^{\top} \in \mathbb{R}^{K}$ is the per-Gaussian weight vector, $w_{i,k} \in [0,1]$ and $\sum_{k} w_{i,k} = 1$, and $K$ is the number of palette colors. To capture view-dependent effects such as specular highlights, these methods decompose each Gaussian's color into diffuse and specular components $\mathbf{c}_i(\theta, \phi) = \mathbf{c}_i^d + \mathbf{c}_i^s(\theta, \phi)$, where the specular component $\mathbf{c}_i^s(\theta, \phi)$ is represented using low-order spherical harmonics~\cite{PaletteGaussian} or neural offsets~\cite{RecolorGaussian}. The weight vector $\mathbf{w}_i$ is either stored directly per-Gaussian or, for compression, obtained by decoding each Gaussian's spatial position through a small Hash-MLP~\cite{muller2022instant}: $\mathbf{w}_i = \Phi(\Psi(\boldsymbol{\mu}_i))$, where $\Psi$ is a multi-resolution hash encoding and $\Phi$ is a decoding network. During training, the palette $\mathbf{P}$, weight representation, and all Gaussian parameters are jointly optimized with sparsity regularization \cite{aksoy2017unmixing} on $\mathbf{w}_i$ to reduce color pollution. Substituting Eq.~\ref{eq:palettegaussian_diffuse} into the 3DGS rendering equation (Eq.~\ref{eq:3dgs_rendering}) and factoring out the palette matrix $\mathbf{P}$, the final pixel color $\mathbf{C}(\mathbf{x}; \theta, \phi)$ becomes:

%\begin{equation} \label{eq:palettegaussian_expanded}
%\mathbf{C}(\mathbf{x}; \theta, \phi) = \sum_{i \in %\mathcal{N}_\mathbf{x}} \left[ \mathbf{w}_i^{\top} %\mathbf{P} + \mathbf{c}_i^s(\theta, \phi) \right] %\alpha_i' \prod_{j=1}^{i-1}(1 - \alpha_j')
%\end{equation}
%Factoring out the palette matrix $\mathbf{P}$, we obtain:

\begin{equation}\label{eq:palettegaussian_factored}
\mathbf{C}(\mathbf{x}; \theta, \phi) = \underbrace{\left[\sum_{i \in \mathcal{N}_\mathbf{x}} \mathbf{w}_i^{\top} \alpha_i' \prod_{j=1}^{i-1}(1 - \alpha_j') \right] \mathbf{P}}_{\text{diffuse: weighted palette}}  + \underbrace{\sum_{i \in \mathcal{N}_\mathbf{x}} \mathbf{c}_i^s(\theta, \phi) \alpha_i' \prod_{j=1}^{i-1}(1 - \alpha_j')}_{\text{specular}}
\end{equation}

This formulation reveals two fundamental limitations of primitive-space decomposition. First, although PaletteGaussian does not explicitly splat weights, pixel-level palette weights (the first term of Eq.~\ref{eq:palettegaussian_factored}) are alpha-blended per-Gaussian weights $\mathbf{w}_i$. Even when each $\mathbf{w}_i$ is sparse and sums to one, blending many Gaussians at a pixel breaks sparsity, producing nonzero weights across most palette colors and causing edits to bleed into unintended regions (Fig.~\ref{fig:sparsity-compare-flip}). In regions of sparse or low-opacity coverage, Gaussian falloff causes pixel weights to sum to less than one, with the residual $(1-\sum_j w_j)$ contributing black and diluting editing control. Second, because the palette $\mathbf{P}$ affects only the diffuse term while specular highlights remain independent (Eq.~\ref{eq:palettegaussian_factored}), enforcing pixel-level color constraints is ill-posed: optimizing the palette alone causes severe bleeding; adding regularization still bleeds, converges slowly (over 100 seconds), and drives the palette to extreme values. Alternatively, jointly optimizing the palette and specular spherical harmonics converges faster but still bleeds, as spherical harmonics encode global view-dependent effects with no clear strategy to impose local regularization (Fig.~\ref{fig:pixel-editing-compare}).

\section{Method}
Our goal is to enable real-time, high-quality palette-based editing for 3DGS scenes with three essential capabilities: direct palette manipulation, per-palette tone curves for independent lightness adjustment, and accurate pixel-level color constraints that propagate consistently across novel views (Fig.~\ref{fig:teaser}). To achieve this, we introduce a view-space palette decomposition framework that addresses the fundamental limitations (\S\ref{sec:motivation}) of primitive-space approaches. Figure~\ref{fig:pipeline} shows our training pipeline.

Our key insight is to splat palette weights instead of colors, enabling palette decomposition directly in view space where editing occurs.
% \yotam{Previous sentence is awkward.}
This framework requires three design choices: (1) a view-dependent weight representation enabling direct regularization in view space, (2) a color space separating lightness from chromaticity for per-palette tone curves, and (3) a new sparsity regularization to enforce sparse, consistent decompositions.
\yotam{Maybe the rest of this paragraph should be moved to the end of the previous paragraph, or just before the ``To achieve this''?}
Building upon ColorfulCurves~\cite{chao2023colorfulcurves}, we extend their palette-based editing paradigm from 2D images to 3D scenes. Like ColorfulCurves, lightness is handled separately from \textit{ab}-space palettes and remapped via per-palette tone curves (\S\ref{sec:editing}). Editing occurs by modifying palette colors or tone curves while keeping decomposition weights fixed, enabling real-time interaction.
% \yotam{If nothing else, you want forward references like ``(\S4.5)'' in this text.}

% \TODO{A semi-technical problem overview of the method. Maybe a problem statement? Or maybe that could be in the Background section?}
% \yotam{A sentence here saying that, like ColorfulCurves, lightness is handled separately from the $ab$-space palettes; they are remapped according to a per-palette tone curve. Or maybe the thing to say is that: Editing is based on changing palette colors while keeping the same weights or remapping lightness with weighted combinations of per-palette tone curves.}

\subsection{View-Space Palette Decomposition}

Each Gaussian retains several standard 3DGS parameters (position, covariance, opacity \S\ref{sec:background}). However, instead of encoding view-dependent colors, we use spherical harmonics (SH) to encode view-dependent palette weights $\mathbf{w}_i \in \mathbb{R}^K$. By making weights view-dependent, we can splat different weight distributions for different viewpoints, capturing view-dependent appearance.
%\deleted{while enabling optimization in the space where users edit}.
For a given view direction $(\theta, \phi)$, the weight vector is
% \begin{equation} \label{eq:colorfulgau_view_weight}
$
\mathbf{w}_i(\theta, \phi) = \sum_{l=0}^{L} \sum_{m=-l}^{l} \allowbreak f_{l,m} Y_l^m(\theta, \phi),
$
% \end{equation}
where $f_{l,m} \in \mathbb{R}^K$ are learned SH coefficients and $Y_l^m$ are the SH basis functions. Following the standard 3DGS rasterization pipeline, we splat these weight vectors to obtain pixel-level palette weights
% \begin{equation}\label{eq:colorfulgau_rendering}
$
\mathbf{W}(\mathbf{x}; \theta, \phi) = \sum_{i \in \mathcal{N}_\mathbf{x}} \mathbf{w}_i(\theta, \phi) \alpha_i' \prod_{j=1}^{i-1}(1 - \alpha_j'),
$
% \end{equation}
where $\mathcal{N}_\mathbf{x}$ is the set of Gaussians affecting pixel $\mathbf{x}$ sorted by depth, and $\alpha_i'$ includes the Gaussian's spatial falloff (Eq.~\ref{eq:3dgs_rendering}).
% \yotam{We say Gaussians everywhere instead of splats or Gaussian splats. Which should we say?}

Unlike primitive-space methods where pixel weights are formed indirectly and do not sum to one (Sec.~\ref{sec:motivation}), our formulation directly produces $\mathbf{W}(\mathbf{x}; \theta, \phi) \in \mathbb{R}^K$ in view space, enabling effective regularization where editing occurs. To ensure convex combinations, we apply softmax normalization:
\begin{equation}\label{eq:softmax_weights_ours}
\widetilde{\mathbf{W}}(\mathbf{x}; \theta, \phi) = \text{softmax}(\mathbf{W}(\mathbf{x}; \theta, \phi)).
\end{equation}
Note that palette weights represent a \emph{color basis decomposition}, not physical radiance contributions. Normalization via softmax ensures interpretable, editable color mixing coefficients regardless of scene coverage.
% \yotam{This would change if we use the alphas sum as the opacity.}
Combined with our sparsity regularization (Sec.~\ref{sec:sparsity_loss}), this yields sparse, consistent decompositions.

Following ColorfulCurves, we adopt CIELAB color space to separate lightness from chromaticity, enabling per-palette tone curves. We define a palette consisting of $K-1$ trainable vertices $\mathbf{p}_1, \ldots, \mathbf{p}_{K-1}$ in \textit{ab}-space. Since all greys in \textit{ab}-space map to the achromatic center and are unlikely to emerge as convex hull vertices during optimization, we explicitly add a fixed grey $\mathbf{z} = (0.5, 0.5)$, where coordinates lie in a normalized\footnote{We normalize the CIELAB color space to $[0, 1]$ for all channels: $L \in [0, 100] \to [0, 1]$, $a \in [-128, 127] \to [0, 1]$, $b \in [-128, 127] \to [0, 1]$. This normalization is used throughout our method for palette representation and SSIM computation.
We store these $ab$ parameters in polar coordinates using clockwise angle offsets and radii $(\Delta \theta, r)$ to avoid explicitly sorting vertices which would break the gradient flow in Sec.~\ref{sec:sparsity_loss}.} \textit{ab}-space range $[0,1]$. 
The complete palette $\mathbf{P} = [\mathbf{z}, \mathbf{p}_1, \ldots, \mathbf{p}_{K-1}] \in \mathbb{R}^{K \times 2}$ thus has $K$ total colors: one fixed achromatic center (grey) and $K-1$ trainable chromatic vertices forming a polygon. This structure ensures direct control over both chromatic and achromatic regions while keeping the palette compact.
Lightness is encoded using separate spherical harmonics per Gaussian. This parallel structure—weights via SH, lightness via SH—ensures both can capture view-dependent effects and enable consistent novel view synthesis. Each Gaussian's view-dependent lightness is computed as:
% \vspace{-1ex}
% \begin{equation}
$
L_i(\theta, \phi) = \sum_{l,m} g_{l,m} Y_l^m(\theta, \phi),
$
% \vspace{-1ex}
% \end{equation}
where $g_{l,m}$ are learned spherical harmonic coefficients. Per-pixel lightness is obtained by splatting these values:
\begin{equation}
L'(\mathbf{x}; \theta, \phi) = \sum_{i \in \mathcal{N}_\mathbf{x}} L_i(\theta, \phi) \alpha_i' \prod_{j=1}^{i-1}(1 - \alpha_j')
\end{equation}
We then apply a sigmoid to ensure the splatted lightness lies in the valid lightness range
% \begin{equation}
$
L(\mathbf{x}; \theta, \phi) = \sigma(L'(\mathbf{x}; \theta, \phi)),
$
% \end{equation}
where $\sigma$ is the sigmoid function ensuring $L(\mathbf{x}; \theta, \phi) \in [0, 1]$. The final \emph{unedited} pixel color in \textit{Lab}-space is:
%\begin{equation}\label{eq:colorfulgau_lab}
% \mathbf{C}_{\textit{Lab}}(\mathbf{x}; \theta, \phi) = \begin{bmatrix} L(\mathbf{x}; \theta, \phi) \\ \widetilde{\mathbf{W}}(\mathbf{x}; \theta, \phi)^\top \mathbf{P} \end{bmatrix}
% \end{equation}
$\mathbf{C}_{\textit{Lab}}(\mathbf{x}; \theta, \phi) = [ L(\mathbf{x}; \theta, \phi),\allowbreak \widetilde{\mathbf{W}}(\mathbf{x}; \theta, \phi)^\top \mathbf{P} ]^\top$,
where chromaticity $\widetilde{\mathbf{W}}(\mathbf{x}; \theta, \phi)^\top \mathbf{P}$ is the weighted combination of palette colors. This decomposition is key to our editing capabilities (Sec.~\ref{sec:editing}): users can modify $\mathbf{P}$ for color changes, apply per-palette tone curves to manipulate $L(\mathbf{x}; \theta, \phi)$, or specify pixel-level constraints on $\mathbf{C}_{\textit{Lab}}(\mathbf{x}; \theta, \phi)$ that account for both diffuse (palette-controlled) and view-dependent (lightness SH-controlled) components.
\yotam{Is it fair to say palette-controlled colors are diffuse (view independent)? Weights are view dependent.}
This addresses both limitations identified in Section~\ref{sec:motivation}: sparse view-space decompositions and comprehensive editing controls previously unavailable in 3DGS palette methods.

\subsection{Sparsity Regularization}
\label{sec:sparsity_loss}
\begin{figure}[hbt!]
\includegraphics[width=\linewidth,scale=1]{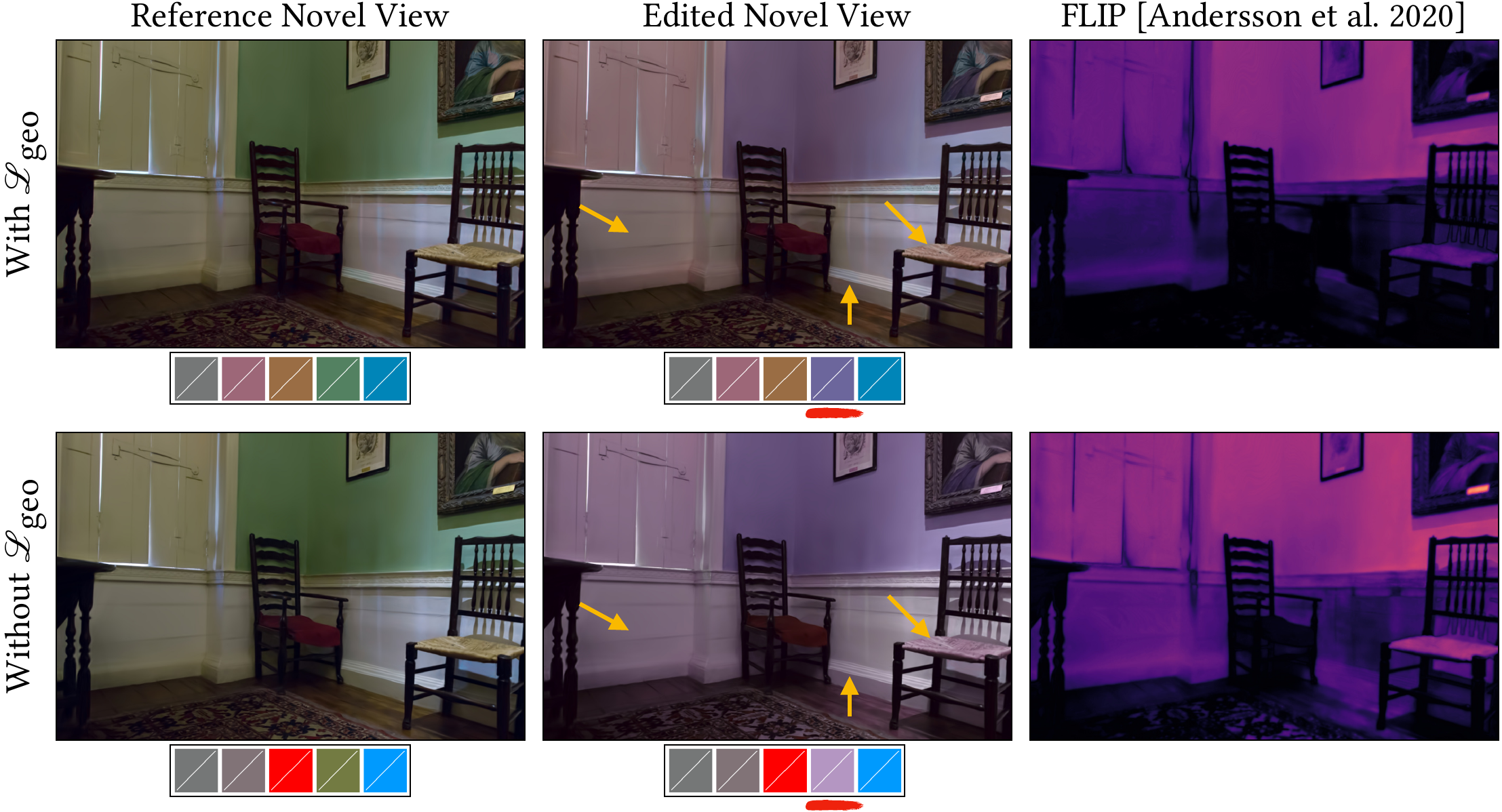}
\caption{Effect of the geometric sparsity loss $\mathcal{L}_{\text{geo}}$. With $\mathcal{L}_{\text{geo}}$, palette weights are consistent and edits remain localized; without, inconsistent decompositions cause visible color bleeding (yellow arrows and FLIP \cite{andersson2020flip}). (\textit{DrJohnson} from Deep Blending~\cite{hedman2018deep}.)}
\label{fig:geometric-sparsity-ablation}
\end{figure}

To prevent color bleeding and unintuitive changes when editing, palette weights should be sparse and perceptually similar colors should have consistent weights (Figs.~\ref{fig:geometric-sparsity-ablation} and \ref{fig:sparsity-compare-flip}).
However, naively optimizing weights to reconstruct a target chroma along with a sparsity term is under-constrained and admits multiple equally sparse decompositions.
For example, \citet{aksoy2017unmixing}'s sparsity $\frac{\sum w_i}{\sum w_i^2}-1$ minimizes the number of non-zero weights per pixel\footnote{\citet{aksoy2017unmixing}'s sparsity is minimized when weights are 0 or 1. Since $w_i \in [0,1]$ and $\sum w_i=1$, it encourages one-hot solutions.} but cannot enforce consistent sparsity patterns for different pixels with similar colors.
%
% While our view-space formulation enables direct optimization of pixel-level weights $\widetilde{\mathbf{W}}(\mathbf{x}; \theta, \phi)$, 
% \yotam{I don't understand the rest of the sentence from here. The first half says we defined our representation. What sparsity regularization do we already know?}
% using sparsity regularization alone in palette-based 3DGS optimization is insufficient for high-quality editing.
% Sparsity measures such as \citet{aksoy2017unmixing}'s minimize the number of nonzero weights but do not enforce consistency across solutions\yotam{what does ``consistency across solutions'' mean?}; when applied in isolation, the resulting optimization is underconstrained and admits multiple equally sparse decompositions\yotam{I don't follow this sentence.}. Consequently, perceptually similar colors may receive inconsistent palette assignments, causing palette edits to bleed into unintended regions (Figs.~\ref{fig:geometric-sparsity-ablation} and~\ref{fig:sparsity-compare-flip}).
%
To address this, we introduce a geometric consistency loss that encourages similar colors to share similar decompositions, combined with sparsity regularization to encourage one-hot distributions.

\paragraph{Geometric consistency.} Our geometric consistency loss is based on a differential formulation of \citet{chao2023colorfulcurves}'s geometric palette weights.
Our loss is expressed in terms of target weights rather than chroma.
% We address this by via a geometric consistency loss encouraging similar colors to share similar decompositions, combined with sparsity regularization for one-hot distributions. 
% Rather than optimizing weights to reconstruct a chroma target directly (admitting infinitely many solutions)
We compute geometric target weights as the inverse barycentric coordinates of the ground truth chroma
in the current palette tessellated into color wedges.
These ground truth weights provide a unique decomposition where perceptually similar colors receive similar weights.
Like ColorfulCurves and prior linear palette methods~\cite{tan2018efficient}, our geometric formulation enforces decompositions to consistently include the grey component, ensuring independent control of chromatic colors without affecting achromatic regions.

Given the palette $\mathbf{P} = [\mathbf{z}, \mathbf{p}_1, \ldots, \mathbf{p}_{K-1}] \in \mathbb{R}^{K \times 2}$ with fixed grey $\mathbf{z} = (0.5, 0.5)$ and $K-1$ trainable chromatic vertices, and a pixel's ground truth chromaticity $\mathbf{c}_{ab}(\mathbf{x}; \theta, \phi)$, we compute target weights $\mathbf{W}_{\text{bary}}(\mathbf{x}; \theta, \phi) \in \mathbb{R}^K$ via soft barycentric assignment. We divide the polygon into $K-1$ triangular wedges, where wedge $i$ is formed by $(\mathbf{z}, \mathbf{p}_i, \mathbf{p}_{i+1})$ with wraparound.
% \revised{We parameterize the palette in polar coordinates using clockwise angle offsets and radii $(\Delta \theta, r)$ to avoid explicit sorting of vertices that would break gradient flow.} \ted{should we also say we do re-parameterization trick for palette in polar coordinates with clockwise offset angles as parameter so that we don't need to do sorting?}.
For each wedge, we compute barycentric coordinates $(\alpha_i, \beta_i, \gamma_i)$ with respect to the three vertices.
% Let $\mathbf{v}_0 = \mathbf{p}_i - \mathbf{z}$, $\mathbf{v}_1 = \mathbf{p}_{i+1} - \mathbf{z}$, and $\mathbf{v}_2 = \mathbf{c}_{ab}(\mathbf{x}; \theta, \phi) - \mathbf{z}$. The barycentric coordinates are:
% \yotam{What follows is just the standard formula for computing the barycentric coordinates $\alpha, \beta, \gamma$ for the ground truth chromaticity $\mathbf{c}_{ab}$, right? I think we can skip these equations for space. OK. I am planning some edits in this section. I also want to change the opening a bit.} \ted{Yes, it is just standard barycentric computation. We can skip. OK! Thank you!}
% \begin{align}
% d_{00} &= \mathbf{v}_0 \cdot \mathbf{v}_0, \quad d_{01} = \mathbf{v}_0 \cdot \mathbf{v}_1, \quad d_{11} = \mathbf{v}_1 \cdot \mathbf{v}_1 \\
% d_{20} &= \mathbf{v}_2 \cdot \mathbf{v}_0, \quad d_{21} = \mathbf{v}_2 \cdot \mathbf{v}_1 \\
% \beta_i &= \frac{d_{11} d_{20} - d_{01} d_{21}}{d_{00} d_{11} - d_{01}^2 + \epsilon}, \quad
% \gamma_i = \frac{d_{00} d_{21} - d_{01} d_{20}}{d_{00} d_{11} - d_{01}^2 + \epsilon} \\
% \alpha_i &= 1 - \beta_i - \gamma_i
% \end{align}
% where $\epsilon = 10^{-9}$ ensures numerical stability.
These coordinates define raw per-wedge weights, where $\alpha_i$ corresponds to the center, $\beta_i$ to vertex $\mathbf{p}_i$, and $\gamma_i$ to vertex $\mathbf{p}_{i+1}$. To handle pixels that may lie outside strict triangular regions or near boundaries, we project each set of barycentric coordinates to a valid simplex via softplus normalization, 
$
\widetilde{\alpha}_i = \frac{\text{softplus}(\alpha_i)}{\text{softplus}(\alpha_i) + \text{softplus}(\beta_i) + \text{softplus}(\gamma_i)},
$
and similarly for $\widetilde{\beta}_i$ and $\widetilde{\gamma}_i$. We reconstruct the chromaticity from each wedge's projected coordinates as $\mathbf{c}_i^{\text{recon}} = \widetilde{\alpha}_i \mathbf{z} + \widetilde{\beta}_i \mathbf{p}_i + \widetilde{\gamma}_i \mathbf{p}_{i+1}$, and measure the reconstruction error $e_i = \|\mathbf{c}_i^{\text{recon}} - \mathbf{c}_{ab}(\mathbf{x}; \theta, \phi)\|_2^2$ for each wedge.
We then differentiably \emph{select} the best wedge
% for a pixel
% by performing soft wedge selection
via softmax over reconstruction errors: 
\begin{equation}
\pi_i = \frac{\exp(-e_i / \tau)}{\sum_{j=1}^{K-1} \exp(-e_j / \tau)}
\end{equation}
where $\tau = 10^{-7}$ is a temperature parameter. This assigns higher probability to wedges with lower reconstruction error. The final target weights are a soft mixture of raw barycentric weights across all wedges: $\mathbf{W}_{\text{bary}}(\mathbf{x}; \theta, \phi) = \sum_{i=1}^{K-1} \pi_i \mathbf{w}_i^{\text{raw}}$, where $\mathbf{w}_i^{\text{raw}} \in \mathbb{R}^K$ is constructed from wedge $i$'s barycentric coordinates. This entire computation is fully differentiable, allowing gradients to flow back to the palette vertices during optimization.

We enforce that our splatted normalized weights $\widetilde{\mathbf{W}}(\mathbf{x}; \theta, \phi)$
%$\mathbf{W}_{\text{gt}}(\mathbf{x}; \theta, \phi)$ 
(Eq.~\ref{eq:softmax_weights_ours}) match these geometric targets via:
\begin{equation}\label{eq:geometric_loss}
\mathcal{L}_{\text{geo}} = \frac{1}{HW} \sum_{\mathbf{x}} \|\widetilde{\mathbf{W}}(\mathbf{x}; \theta, \phi) - \mathbf{W}_{\text{bary}}(\mathbf{x}; \theta, \phi)\|_2^2
\end{equation}
% \yotam{The tilde and hat are visually similar. Maybe change hat to $W^\textit{geo}$ or $W_\textit{geo}$?}
where $H$ and $W$ are the image resolution. 
Note that $\mathbf{W}_{\text{bary}}(\mathbf{x}; \theta, \phi)$ may contain negative weights when a pixel's chromaticity lies outside the current palette. While negative weights are not interpretable for color editing, they serve a crucial role. Since $\widetilde{\mathbf{W}}(\mathbf{x}; \theta, \phi)$ from softmax is strictly positive and sums to one, minimizing $\|\widetilde{\mathbf{W}}(\mathbf{x}; \theta, \phi) - \mathbf{W}_{\text{bary}}(\mathbf{x}; \theta, \phi)\|^2_2$ when target weights are negative encourages palette expansion, ensuring the palette grows to encompass the full color distribution. This loss also ensures perceptually similar colors receive similar decompositions, preventing inconsistent sparsity patterns that cause editing artifacts (Figs.~\ref{fig:geometric-sparsity-ablation} and \ref{fig:sparsity-compare-flip}). 

\paragraph{Sparsity.} To encourage one-hot weight distributions among chromatic colors, we apply the sparsity measure from~\citet{aksoy2017unmixing} only to the chromatic palette weights, excluding grey.
Including grey would bias the solution toward grey-dominated solutions, since minimum sparsity can be achieved by simultaneously increasing grey weights while expanding the chromatic palette vertices toward the gamut boundary, losing representativeness.
We do not renormalize the chromatic weights and then measure sparsity, as this would allow
% \yotam{What are renormalized chromatic weights? It means: renormalizing the non-grey weights to sum to 1 and then measuring sparsity.}
the optimizer to exploit scale-invariance to achieve both low sparsity (through small but one-hot renormalized weights) and low reconstruction error (via large palettes with grey-dominated decompositions). Therefore, we instead compute:
\begin{equation}\label{eq:sparsity_loss}
\mathcal{L}_{\text{sparse}} = \frac{1}{HW} \sum_{\mathbf{x}} \left( \frac{\sum_{k=2}^{K} \widetilde{W}_k(\mathbf{x}; \theta, \phi)}{\sum_{k=2}^{K} \widetilde{W}_k(\mathbf{x}; \theta, \phi)^2 + \epsilon} - 1 \right)
\end{equation}
where $\epsilon = 10^{-8}$. This formulation penalizes both non-sparse chromatic weight patterns and grey-dominated solutions (where chromatic weights become uniformly small), encouraging representative chromatic decompositions.
% \yotam{Is this different from Aksoy?}

%%% Yotam continue here

\subsection{Palette Regularization}
\label{sec:palette_reg}
The geometric consistency loss (Sec.~\ref{sec:sparsity_loss}) encourages the palette to expand to cover chromaticity points that lie outside its current extent while maintaining consistent sparsity patterns. However, unconstrained expansion would produce overly large, unrepresentative palettes that fail to capture the scene's dominant color distribution (Fig.~\ref{fig:grey-penalization-ablation}). We introduce two regularization terms to maintain palette compactness and prevent degeneracy. First, we penalize the grey weight with a loss $\mathcal{L}_\text{grey}$ to encourage palette compactness:
\vspace{-.25ex}
\begin{equation}\label{eq:grey_loss}
\mathcal{L}_{\text{grey}} = \frac{1}{HW} \sum_{\mathbf{x}} \widetilde{W}_1(\mathbf{x}; \theta, \phi)
\vspace{-.25ex}
\end{equation}
where $H$ and $W$ are the image dimensions, and  $\widetilde{W}_1$ is the grey weight (first component of $\widetilde{\mathbf{W}}$).
Since the fixed grey contributes to all colors via barycentric coordinates,
minimizing the grey contribution increases the contribution of the other colors in the palette.
Since grey
$\mathbf{z} = (0.5, 0.5)$ lies at the origin of our normalized \textit{ab}-space,
% \yotam{What does ``our normalized'' mean?} It's from earlier.
this shrinks the palette toward the center.
This loss balances the geometric loss's expansion pressure: when combined, the two losses encourage the palette to grow just enough to encompass the scene's color distribution while remaining compact.

Second, to prevent palette vertices from collapsing into each other (which would reduce the palette's representational capacity), we enforce a minimum separation between chromatic vertices:
\vspace{-.25ex}
\begin{equation}\label{eq:palette_sep_loss}
\mathcal{L}_{\text{palette}} = \sum_{i=2}^{K} \sum_{j=i+1}^{K} \max(0, d_{\min} - \|\mathbf{p}_i - \mathbf{p}_j\|_2)
\vspace{-.25ex}
\end{equation}
where $d_{\min} = 0.1$ is the minimum allowed distance in normalized \textit{ab}-space.
This loss is only active
% \yotam{Is ``activate'' referring to the $\max$?} \ted{yes}
when vertices become too close, preventing degeneracy while providing the optimizer the freedom to position vertices optimally.

\subsection{Optimization}
\label{sec:optimization}

We optimize our model following the standard 3DGS training pipeline \cite{kerbl20233d} with adaptive density control, but replace the color prediction with our view-space palette decomposition and additional regularization losses. 
We supervise lightness reconstruction using a combination of L1 and SSIM terms:
\vspace{-.25ex}
\begin{equation}\label{eq:lightness_loss}
\begin{split}
\mathcal{L}_{\text{lightness}} = &\frac{\lambda_l}{HW} \cdot \sum_\mathbf{x}\| L(\mathbf{x}; \theta, \phi) - L^{\text{gt}}(\mathbf{x}; \theta, \phi)\|_2^2 \\
\vspace{-.25ex}
&+ (1-\lambda_l) \cdot [1 - \text{SSIM}\left(L(\mathbf{x}; \theta, \phi), L^{\text{gt}}(\mathbf{x}; \theta, \phi)\right)]
\end{split}
\vspace{-.25ex}
\end{equation}
where $L^{\text{gt}}$ is the ground truth lightness channel in Lab-space
% \yotam{Lab or LAB?}\ted{Let's do Lab-space and full name CIE LAB space.}
and SSIM measures structural similarity. We apply reconstruction loss only on the lightness channel; chromaticity is implicitly supervised through the geometric consistency loss (Eq.~\ref{eq:geometric_loss}), which enforces that predicted weights match barycentric coordinates computed from ground truth \textit{ab} values. The overall training objective optimizes all Gaussian parameters $\Theta$ and the palette $\mathbf{P}$ to minimize 
% \begin{equation}\label{eq:total_loss}
% \begin{aligned}
% \mathcal{L} = &\mathcal{L}_{\text{lightness}} + \mathcal{L}_{\text{geo}} + \lambda_{\text{sparse}} \mathcal{L}_{\text{sparse}} \\
% &+ \lambda_{\text{grey}} \mathcal{L}_{\text{grey}} + \lambda_{\text{palette}} \mathcal{L}_{\text{palette}}
% \end{aligned}
% \end{equation}
$\mathcal{L} = \mathcal{L}_{\text{lightness}} + \mathcal{L}_{\text{geo}} + \lambda_{\text{sparse}} \mathcal{L}_{\text{sparse}} + \lambda_{\text{grey}} \mathcal{L}_{\text{grey}} + \lambda_{\text{palette}} \mathcal{L}_{\text{palette}}$,
where $\Theta = \{\boldsymbol{\mu}_i, \boldsymbol{\Sigma}_i, \alpha_i, \{f_{l,m}\}_i, \{g_{l,m}\}_i\}$ represents all Gaussian parameters (positions, covariances, opacities, and SH coefficients for weights and lightness), and $\mathbf{P} = [\mathbf{z}, \mathbf{p}_1, \ldots, \mathbf{p}_{K-1}]$ is the palette with fixed grey $\mathbf{z}$ and trainable chromatic vertices. Here, $\mathcal{L}_{\text{geo}}$ is the geometric consistency loss (Eq.~\ref{eq:geometric_loss}), $\mathcal{L}_{\text{sparse}}$ is the sparsity loss (Eq.~\ref{eq:sparsity_loss}), $\mathcal{L}_{\text{grey}}$ penalizes grey dominance (Eq.~\ref{eq:grey_loss}), and $\mathcal{L}_{\text{palette}}$ enforces vertex separation (Eq.~\ref{eq:palette_sep_loss}). We randomly initialize palette vertices in normalized \textit{ab}-space and all Gaussian parameters $\Theta$ following standard 3DGS~\cite{kerbl20233d}. While palette extraction methods~\cite{tan2018efficient} can compute initial palettes from training images, we found random initialization produces comparable results, suggesting our geometric consistency loss effectively guides palette optimization regardless of initialization. Complete training hyperparameters and implementation details are provided in Sec.~\ref{sec:results}.

\subsection{Real-Time Editing Controls}
\label{sec:editing}

Our view-space decomposition naturally extends ColorfulCurves from 2D to 3D. Once trained, users can edit the scene in real-time through three types of controls: palette manipulation, per-palette tone curves, and pixel-level constraints.
% Unlike primitive-space methods where palette edits require re-optimizing view-dependent components to achieve more balanced editing (Fig.~\ref{fig:pixel-editing-compare}),
% \yotam{What is this referring to? Citation?}
% our formulation allows direct editing through sparse optimization on the palette and curves alone, while keeping all Gaussian parameters fixed.
Following ColorfulCurves, given $K$ palette colors, we define
% monotonic \yotam{They don't have to be monotonic, do they?} \ted{right. they are identity maps in the beginning.}
curve functions $\mathcal{F} = \{f_1, \ldots, f_K\}$, where each $f_i: [0,1] \rightarrow [0,1]$ controls the lightness mapping for palette color $\mathbf{p}_i$. To preserve the original black and white points, we enforce $f_i(0) = 0$ and $f_i(1) = 1$. Let $L_0(\mathbf{x}; \theta, \phi)$ denote the initially splatted lightness from training for pixel $\mathbf{x}$ at camera viewpoint $(\theta, \phi)$. The edited lightness is computed as:
\vspace{-.5ex}
\begin{equation}\label{eq:tone_curves}
C_L(\mathbf{x}; \theta, \phi) = \sum_{i=1}^{K} \widetilde{\mathbf{U}}_i(\mathbf{x}; \theta, \phi) \cdot f_i(L_0(\mathbf{x}; \theta, \phi))
\vspace{-.5ex}
\end{equation}
where $\widetilde{\mathbf{U}}$ is obtained by scaling the grey weight (index 1) of 
$\widetilde{\mathbf{W}}$ (Eq.~\ref{eq:softmax_weights_ours}) by 0.01 and renormalizing to sum to one.
This ensures that edits made to lightness will be dominated by chromatic palette colors.
% \yotam{What are these renormalized weights for?}
Initially, each $f_i$ is the identity map. The final rendered color combines edited lightness with palette-based chromaticity:
% \begin{equation}\label{eq:colorfulgau_lab_render_edit}
% \mathbf{C}_{\text{Lab}}(\mathbf{x}; \theta, \phi) = \begin{bmatrix} C_L(\mathbf{x}; \theta, \phi) \\ \widetilde{\mathbf{W}}(\mathbf{x}; \theta, \phi)^\top \mathbf{P} \end{bmatrix}
% \end{equation}
$\mathbf{C}_{\text{Lab}}(\mathbf{x}; \theta, \phi) = [ C_L(\mathbf{x}; \theta, \phi), \widetilde{\mathbf{W}}(\mathbf{x}; \theta, \phi)^\top \mathbf{P} ]^\top$.

\paragraph{Palette manipulation.}
Users can directly modify palette vertices $\mathbf{p}_i$ in \textit{ab}-space. $\mathbf{C}_{\text{Lab}}$ chromaticity is computed as $\widetilde{\mathbf{W}}(\mathbf{x}; \theta, \phi)^\top \mathbf{P}$, so palette edits immediately propagate to all pixels according to their weights, maintaining multi-view consistency. This enables global recoloring operations.

\paragraph{Per-palette tone curves.}
Users can place control points on individual curves $f_i$ to adjust lightness behavior for specific palette colors. For example, users can 
%brighten reds without affecting blues \TODO{ddd}, or 
add contrast to specific color ranges via S-shaped curves (Figs.~\ref{fig:teaser} and ~\ref{fig:palette-aware-lightnes-compare}), a capability unavailable in prior palette-based 3DGS methods. To ensure smooth lightness transitions, each curve $f_i$ is a biharmonic function obtained variationally by minimizing the squared Laplacian subject to the user's control points and natural boundary conditions~\cite{chao2023colorfulcurves}.
% constrained to interpolate the 
% \yotam{Each curve is a biharmonic function with natural boundary conditions constrained to interpolate the constraints?}

\paragraph{Pixel-level constraints.}
% \todo{Clarify that a pixel-level constraint is a constraint on a point in a rendered view. Equivalent to a constraint along a viewing ray. Radiance? Light field!}
Users can specify the desired color for selected pixels by clicking in any rendered view (Figs.~\ref{fig:teaser} and \ref{fig:pixel-editing-compare}).
% This is equivalent to constraining the radiance along a view-space ray.
% \revised{This is equivalent to constraining the radiance field.}
Following ColorfulCurves, we solve an $L_{2,1}$ optimization for the sparsest change to the palette $\mathbf{P}$ and curves $\{f_i\}$ that satisfy all user constraints (pixel, palette, and tone curve) via a fast alternating optimization (Algorithm 1 in~\cite{chao2023colorfulcurves}).
% \yotam{block coordinate descent? Was one block quadratic?}
A key advantage of our formulation is that pixel constraints only require optimizing the palette and curves, leaving the many Gaussian parameters fixed. This drastically reduces the degrees of freedom and runs in real-time. In contrast, primitive-based methods must re-optimize both palette and view-dependent representations (spherical harmonics \cite{PaletteGaussian} or neural offsets \cite{RecolorGaussian}). 

% \yotam{The next few sentences reads like something from results or evaluation.}
%Primitive optimization prevents real-time interaction and produces visual artifacts. Non-sparse pixel weights cause edits to bleed into unintended regions, and over-regularizing the palette forces the optimizer to compensate through view-dependent components originally intended for specular effects, resulting in unnatural appearance (Figs.~\ref{fig:sparsity-compare-flip}, \ref{fig:pixel-editing-compare}). Our optimization converges in under 0.02 seconds, enabling interactive constraint-driven editing with superior quality.

%%%%%%%%%%%%% Results %%%%%%%%%%%%%
\section{Results and Evaluation} \label{sec:results}
\begin{figure*}[t!]
\includegraphics[width=\linewidth,scale=1]{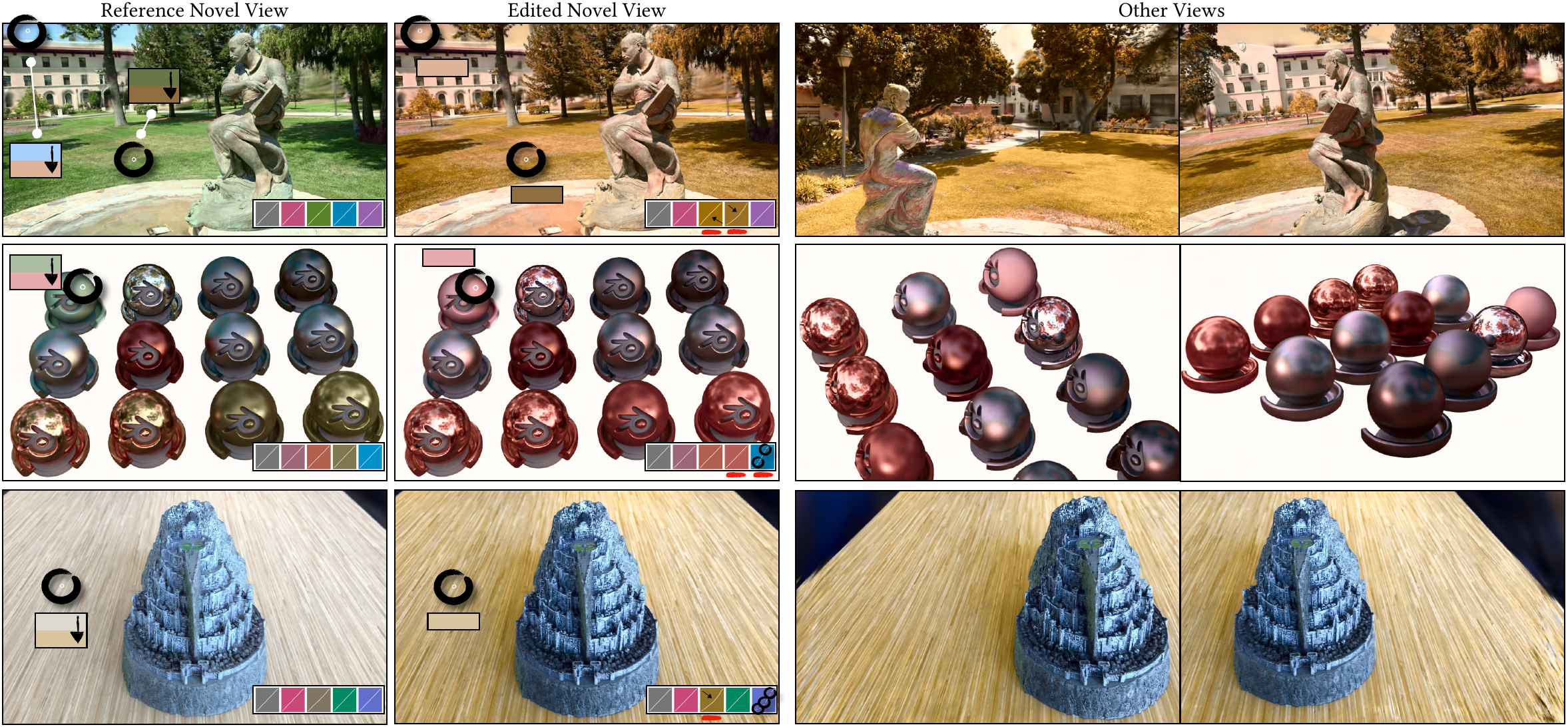}
\caption{\system can edit a variety of scene in real time by placing pixel-level constraints on reference view. Datasets (top to bottom): \textit{Ignatius} from Tanks\&Temple \cite{knapitsch2017tanks}; \textit{Materials} from Blender \cite{mildenhall2021nerf}; \textit{Fortress} from LLFF \cite{mildenhall2019local}.}
\label{fig:gallery1}
\end{figure*}

\begin{figure}[hbt!]
\includegraphics[width=\linewidth,scale=1]{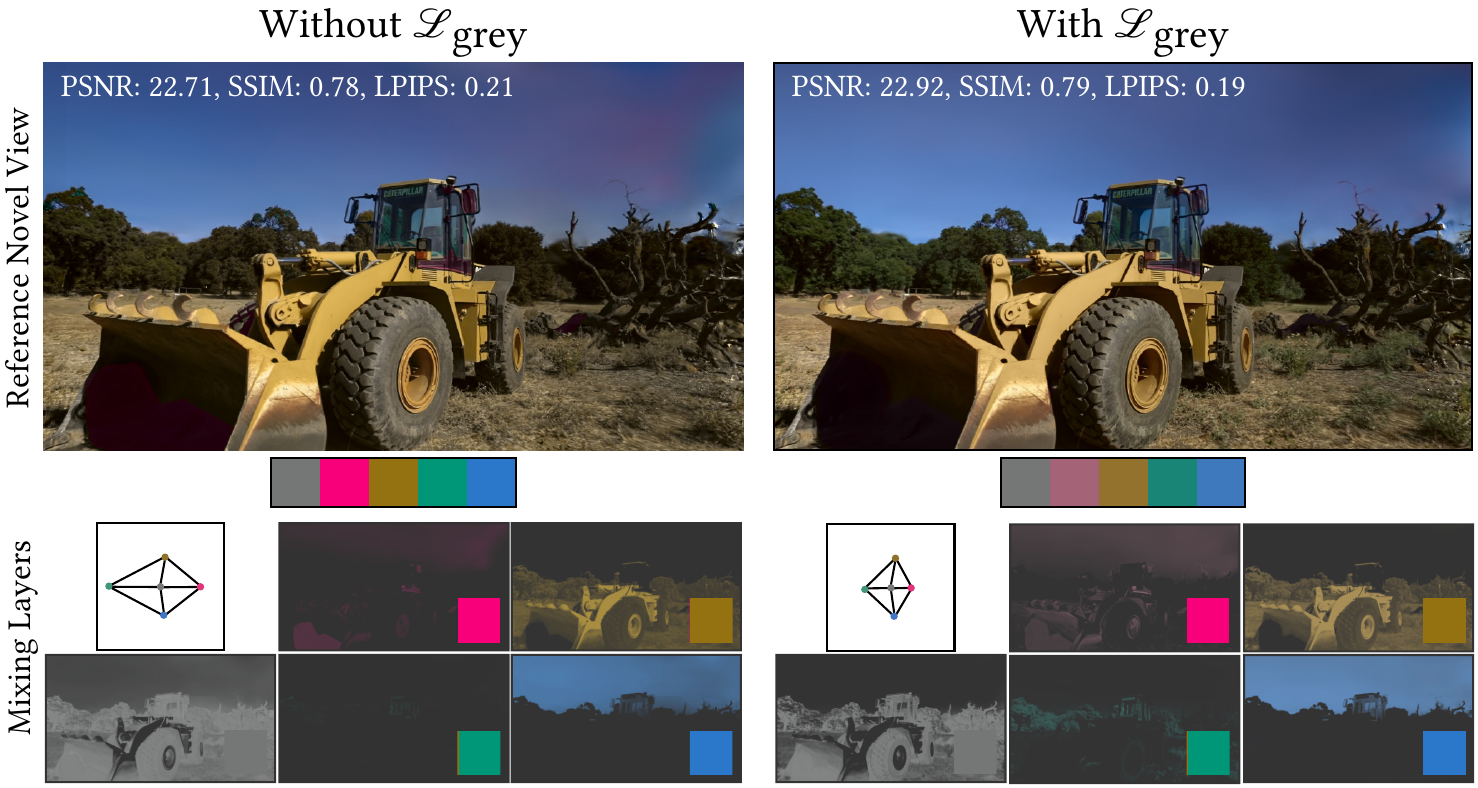}
\caption{Grey penalization ablation. Without $\mathcal{L}_{\text{grey}}$, grey dominates the palette weights, resulting in an unrepresentative palette and reduced editability, as chromatic colors (pink and green) receive negligible weights. Enforcing $\mathcal{L}_{\text{grey}}$ yields a compact palette in which chromatic colors meaningfully control the scene, improving editability without sacrificing reconstruction quality (PSNR, SSIM, and LPIPS are shown inset). The palette plot is shown in \textit{ab}-space with a consistent relative scale. (\textit{Caterpillar} from Tanks\&Temple \cite{knapitsch2017tanks}.)}
\label{fig:grey-penalization-ablation}
\end{figure}

\begin{figure*}[hbt!]
\includegraphics[width=0.9\linewidth]{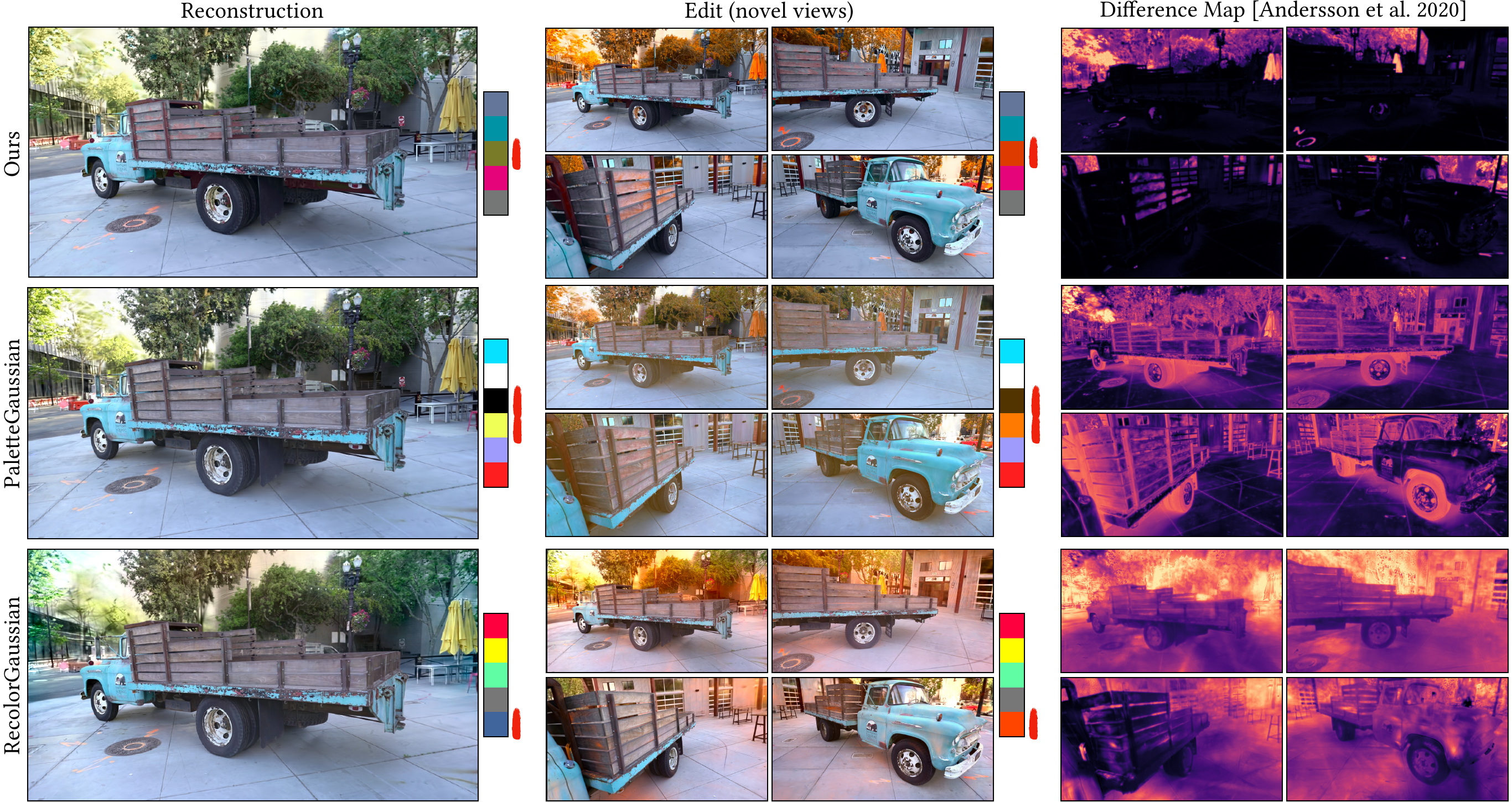}
\caption{Editing sparsity comparison with PaletteGaussian~\cite{PaletteGaussian} and RecolorGaussian~\cite{RecolorGaussian}. Given reconstructed views (left), we perform real-time palette editing to change forest from green to autumn by modifying a single palette color (red underlines, middle column showing novel view edits). Difference maps (right, FLIP~\cite{andersson2020flip}) demonstrate our method produces the most localized edits. PaletteGaussian requires modifying both yellow and black palette colors because forest pixels have inconsistent weight patterns: some forest regions are dominated by yellow weights while others by black weights, causing color bleeding despite sparsity regularization. RecolorGaussian exhibits similar artifacts. Both primitive-space methods suffer from inconsistent sparsity patterns where perceptually similar colors receive different palette decompositions, while our view-space geometric loss enforces consistent decompositions, achieving superior editing locality. \textit{Truck} from Tanks\&Temple \cite{knapitsch2017tanks}.}
\label{fig:sparsity-compare-flip}
\end{figure*}

\begin{figure*}[hbt!]
\includegraphics[width=\linewidth,scale=1]{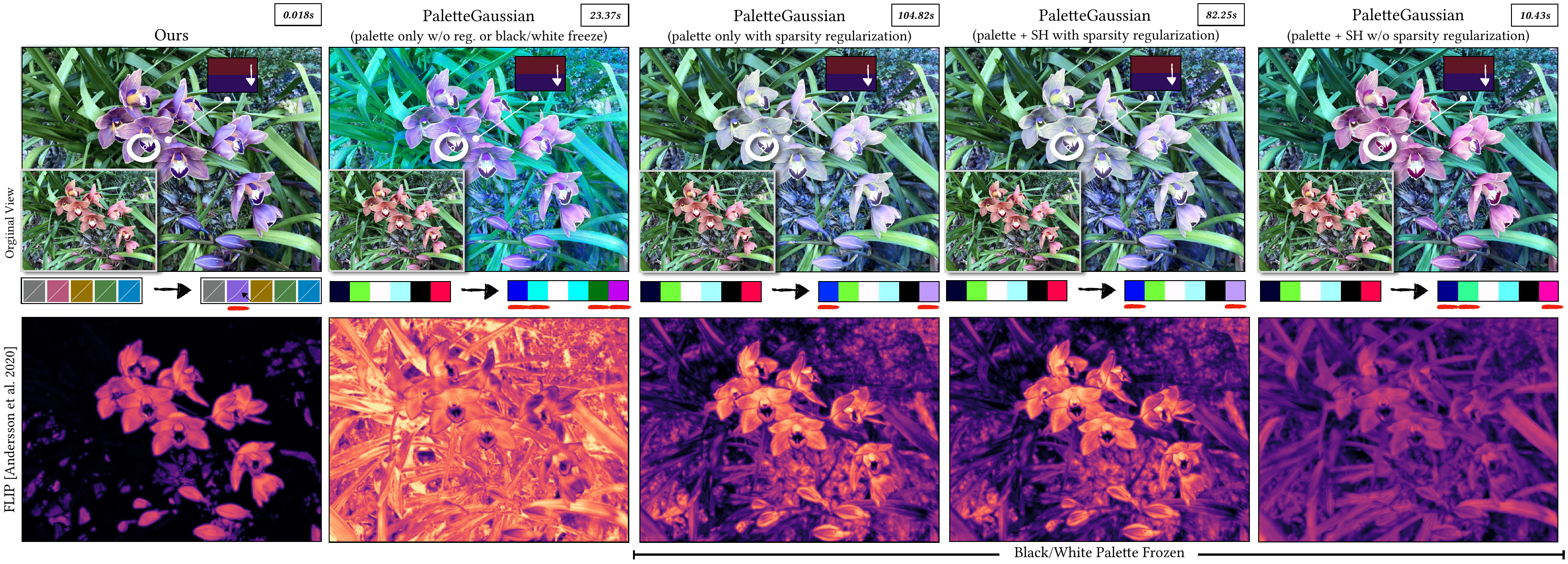}
\caption{Pixel-level editing comparison with PaletteGaussian~\cite{PaletteGaussian} (input views inset bottom-left). Given a user-specified color constraint (pink flowers), our method (leftmost) achieves real-time editing (0.018s) with superior quality. We compare four PaletteGaussian strategies: (1) palette only without regularization or black/white freeze (23.37s), (2) palette only with $L_{2,1}$ sparsity and freeze (104.82s), (3) joint palette and spherical harmonics optimization with sparsity and freeze (Eq.~\ref{eq:palettegaussian_factored}, 82.25s), and (4) joint optimization with freeze but no sparsity (10.43s). Freezing black and white (strategies 2-4, above the black line) prevents whole-scene bleeding, yet even the fastest strategy (10.43s) exhibits significant color bleeding into the background (e.g., green tint). All strategies remain far from real-time. FLIP \cite{andersson2020flip} difference maps (bottom row) show our method produces highly localized edits affecting only the target flowers, while PaletteGaussian suffers from non-sparse pixel-level weights due to primitive-space decomposition. (\textit{Orchids} from LLFF \cite{mildenhall2019local}.)}
\label{fig:pixel-editing-compare}
\end{figure*}

We demonstrate a range of editing capabilities enabled by \system, including palette-based recoloring (Figs.~\ref{fig:teaser}, \ref{fig:grey-penalization-ablation}, and \ref{fig:sparsity-compare-flip}), palette-aware lightness editing (Figs.~\ref{fig:teaser}, \ref{fig:palette-aware-lightnes-compare}, and \ref{fig:gallery1}), and pixel-level color constraints (Figs.~\ref{fig:teaser}, \ref{fig:gallery1}, and \ref{fig:pixel-editing-compare}). All edits are performed in real time and propagate consistently across novel views.
% \todo{need to fix figure ordering. figures should not fit into two pages.}

\paragraph{Implementation and Performance}
We implemented \system in PyTorch with custom CUDA kernels for efficient weight splatting. We use $K=5$ palette colors (one fixed grey plus 4 trainable chromatic colors) and spherical harmonics up to degree 3 for both weight and lightness encoding. For visualization, all palettes are shown at constant lightness $L=50$ in CIELAB color space.
We train each scene for 200--500 epochs using the Adam optimizer~\cite{kingma2014adam}, with each epoch shuffling the training views.
We adopt the standard evaluation protocol of \citet{barron2021mip}, using every 8th image as a held-out test view and training on the remaining images, and report reconstruction quality using PSNR, SSIM, and LPIPS for consistent comparison (Table~\ref{tb:recon-quality}).
In practice, we set $\lambda_{\text{sparse}} = 0.002$, $\lambda_{\text{grey}} = 0.03$ and $\lambda_{\text{palette}} = 0.1$. Training takes approximately 40--60 minutes on an NVIDIA A100-40GB, depending on scene complexity.
On the same GPU, our approach renders at ${\sim}175.74$ FPS at 800$\times$600 resolution on the LLFF dataset. Our real-time editing optimization takes ${\sim}0.02$ seconds and is performed when the user edits (i.e.\ constraints change).
% \yotam{Separate rendering from optimization. Say this leads to real-time editing.}
% We store the optimization result and keep track with any changes been made to the current constraint status, only re-run optimization when is needed.

\begin{table}[t]
\centering
\caption{Reconstruction quality on test views from the LLFF dataset~\cite{mildenhall2019local}. Our method achieves comparable reconstruction quality to prior palette-based methods while enabling view-space decomposition for superior editing capabilities. Baseline results are from \citet{PaletteGaussian}.}
\label{tab:reconstruction}
\begin{tabular}{lccc}
\toprule
Method & PSNR $\uparrow$ & SSIM $\uparrow$ & LPIPS $\downarrow$ \\
\midrule
PaletteNeRF~\cite{kuang2023palettenerf} & 22.46 & 0.70 & 0.21 \\
RecolorNeRF~\cite{gong2023recolornerf} & 24.32 & 0.77 & \textbf{0.13} \\
PaletteGaussian~\cite{PaletteGaussian} & 23.75 & 0.79 & 0.18 \\
\midrule
Ours & \textbf{24.52} & \textbf{0.82} & 0.14 \\
\bottomrule
\end{tabular}
\label{tb:recon-quality}
\end{table}

% \subsection{Comparison}
\paragraph{Comparison}
% \paragraph{Evaluation.} 
We evaluate \system on three aspects: reconstruction fidelity to verify that editability does not degrade novel-view synthesis (Table~\ref{tb:recon-quality}), edit locality under palette modifications (Fig.~\ref{fig:sparsity-compare-flip}), and interactive editing capabilities including palette manipulation, tone curve adjustment, and pixel-level constraints (Figs.~\ref{fig:palette-aware-lightnes-compare} and~\ref{fig:pixel-editing-compare}). We compared against our implementations of PaletteGaussian~\cite{PaletteGaussian}\footnote{We implement explicit per-Gaussian palette weights instead of Hash-MLP compression to provide a more expressive baseline, as the Hash-MLP is primarily a storage optimization.} and RecolorGaussian~\cite{RecolorGaussian}. Figure~\ref{fig:sparsity-compare-flip} demonstrates that primitive-space decomposition produces non-sparse pixel weights after alpha-blending, causing palette edits to bleed into unintended regions even with per-Gaussian sparsity regularization. Figure~\ref{fig:pixel-editing-compare} shows that pixel-level editing in primitive-space methods requires joint optimization of palette and view-dependent representations, taking over 10 seconds while still exhibiting color bleeding. Over-regularizing the palette forces the optimizer to compensate through view-dependent components originally intended for specular effects, resulting in an unnatural appearance. In contrast, our view-space formulation achieves localized edits through sparse optimization over palettes and curves alone, with all Gaussian parameters fixed. Our optimization converges in 0.02 seconds, enabling interactive constraint-driven editing with superior locality and visual quality.
While PaletteGaussian proposes object-level editing through second-stage training with segmentation masks, the masks shown in their work are coarse and lack the soft boundaries needed for natural color compositing in complex regions such as forests (Fig.~\ref{fig:sparsity-compare-flip}), as noted by \citet{chao2023locopalettes}. More critically, even when segmentation succeeds, their editing operates on the same non-sparse pixel weights, fundamentally limiting edit locality and producing color bleeding.
 
%\yot{In contrast, approaches based on} primitive optimization (\cite{PaletteGaussian, RecolorGaussian}) \yot{citations} prevents real-time interaction and produces visual artifacts. Non-sparse pixel weights cause edits to bleed into unintended regions, and over-regularizing the palette forces the optimizer to compensate through view-dependent components originally intended for specular effects, resulting in unnatural appearance (Figs.~\ref{fig:sparsity-compare-flip}, \ref{fig:pixel-editing-compare}). Our optimization converges in under 0.02 seconds, enabling interactive constraint-driven editing with superior quality.

\paragraph{Ablations.} 
Our geometric regularization loss $\mathcal{L}_{\text{geo}}$ enforces consistent sparsity patterns, preventing similar colors from receiving inconsistent palette assignments that cause edits to bleed unpredictably (Fig.~\ref{fig:geometric-sparsity-ablation}). Without this loss, achromatic pixels may be represented as mixtures of complementary chromatic colors rather than through our fixed grey vertex (analogous to the line of greys in RGB space~\cite{tan2018efficient}), producing perceptually prominent tinting artifacts when palette colors are edited. Figure~\ref{fig:grey-penalization-ablation} demonstrates the importance of our grey penalization loss $\mathcal{L}_{\text{grey}}$ for maintaining representative palettes. Removing $\mathcal{L}_{\text{grey}}$ causes the palette to expand with grey dominating the decomposition, leaving chromatic colors with negligible weights and severely limiting editability. Our compact palette with $\mathcal{L}_{\text{grey}}$ maintains comparable reconstruction quality while providing superior editability.

\section{Conclusion}
\label{sec:conclusion}

We have presented \system, a real-time color grading approach for 3D Gaussian Splatting that enables high-quality, multi-view consistent color editing. We accomplished this with a view-space sparse palette decomposition. \system achieves superior editing locality and consistency compared to prior approaches, while maintaining comparable reconstruction fidelity and enabling real-time palette, tone curve, and direct pixel editing. Extensive experiments demonstrate the effectiveness of our approach across a variety of scenes and editing scenarios.

\paragraph{Limitations and Future Work}
Our method has several limitations that suggest directions for future work.
% First, we visualize palettes at constant lightness $L=50$ in \textit{ab}-space for consistency across views. While \citet{chao2023colorfulcurves} visualizes palette colors using weight-averaged lightness from the image, this approach is challenging in multi-view settings where palette appearance would change across viewpoints, making it difficult for users to track editing constraints. Developing view-invariant palette visualizations that better convey lightness information remains an open problem.
Encoding view-dependent weights with spherical harmonics requires approximately 2$\times$ the storage of standard 3DGS for typical palette sizes ($K=5$ chromatic colors plus lightness). More efficient view-dependent representations such as spherical gaussians~\cite{wang2024sg}, spherical beta~\cite{liu2025deformable}, or spherical Voronoi~\cite{di2025spherical} could reduce this overhead while maintaining editability.
Our method does not support RGBA rendering. When renormalizing our splatted weights, we could treat the sum as $\alpha$ for a premultiplied color representation.
Finally, our method does not support semantic or object-level editing constraints. Incorporating segmentation-aware decompositions \cite{chao2023locopalettes,aksoy2018semantic}, demonstrated for NeRFs~\cite{chenkarf}, could enable selective recoloring of specific objects or materials while maintaining our view-space sparsity advantages.

\begin{acks}
% \section*{Acknowledgements}
% None for now.
\end{acks}

\bibliographystyle{ACM-Reference-Format}
\bibliography{bib/merged,bib/gaussian}

%% https://tex.stackexchange.com/questions/45609/is-it-wrong-to-use-clearpage-instead-of-newpage
%% Clearpage eats figures
% \clearpage

\phantom{.}

%%%%%%%% Figure Pages %%%%%%%%%%%

\clearpage

\end{document}